\documentclass[12pt]{article}
\usepackage[centertags]{amsmath}
\usepackage{amsfonts}
\usepackage{amssymb}
\usepackage{amsthm}
\usepackage{newlfont}
\begin{document}

\author{D. M. Walsh\thanks{E-mail
: darraghmw@gmail.com}\\ \small Physics Department,\\
\small University College, Cork, Ireland}

\title{Non-uniqueness in conformal formulations of the Einstein Constraints}
\maketitle
\begin{abstract}
Standard methods in non-linear analysis are used to indicate that
there exists a parabolic branching of solutions of the
Lichnerowicz-York equation with an unscaled source. We also apply
these methods to the extended conformal thin sandwich formulation
and, by assuming that the linearised system develops a kernel
solution for sufficiently large initial data, we reproduce the
parabolic solution curves for the conformal factor, lapse and shift
found numerically by Pfeiffer and York.
\end{abstract}
\section{Introduction}
With the onset of gravitational wave astronomy approaching, it is of
crucial importance to have constructed suitably realistic
theoretical and numerical models of the space-time structure of
gravitational wave sources. This demands that we study the initial
value problem for Einstein's gravitational field equations. These
form a complicated quasi-linear system of partial differential
equations. They naturally split into six evolution equations and
four constraint equations on the initial data. It is the constraint
equations that this work will focus on.

This work was initiated to try to explain the intriguing
non-uniqueness results found numerically by Pfeiffer and York in
\cite{PY1}. They found a parabolic curve of regular solutions for
each of the five unknowns in the extended conformal thin sandwich
formulation of the Einstein constraints. This system has enjoyed
much support amongst numerical relativists but their results show
the existence of two regular solutions for the entire range of
wave amplitude considered. We present a simple local derivation of
these non-uniqueness results.

We first review the important features of the conformal method for
solving the constraints. The four constraint equations of general
relativity are
\begin{equation}\label{ham}
  R(\bar{g}) -\bar{K}^{ij}\bar{K}_{ij}\;+\;\bar{K}^2=16\pi \rho
  \end{equation}
  \begin{equation}\label{mom}
  \bar{\nabla}_i(\bar{K}^{ij}-\bar{g}^{ij}\bar{K})=8\pi  J^j
\end{equation}where $\bar{K}^{ij}$ is the
extrinsic curvature of the space-like initial data slice and
$\bar{K}=tr_{\bar{g}}\bar{K}^{ij}$ and $\rho$ and $J^j$ must
satisfy conservation equations. In the various conformal
formulations an initial 3-metric $g_{ij}$ is chosen which is
conformally related to the physical solution of the constraints
$\bar{g}_{ij}$
\begin{equation}\label{CT1}
\bar{g}_{ij}=\phi^4g_{ij}.
\end{equation}We decompose $\bar{K}^{ij}$ as $\bar{K}^{ij}=\bar{A}^{ij}+\frac{1}{3}\bar{g}^{ij}\bar{K}$
and define the trace-free conformal extrinsic curvature tensor
${A}^{ij}$ according to
\begin{equation}\label{CT2}
\bar{A}^{ij}=\phi^{-10}A^{ij}
\end{equation}which leads to
\begin{equation}
 \bar{A}^{ij}\bar{A}_{ij}=\phi^{-12}{A}^{ij}{A}_{ij},
 \end{equation} where barred objects are defined with respect to
 the physical metric $\bar{g}$ and $\bar{K}=K$.
This property is used in the transformation of the Hamiltonian
constraint (\ref{ham}). Together with the transformation for the
scalar curvature given by
\begin{equation}
R(\bar{g})\phi^5=R(g)\phi-8{\nabla}^2\phi
\end{equation} it yields the Lichnerowicz-York (LY) equation for
$\phi$:
\begin{equation}\label{LY}
{\nabla}^2\phi-\frac{{R}(g)\phi}{8}=-\frac{1}{8}{A}_{ij}{A}^{ij}\phi^{-7}+
\frac{K^2}{12}\phi^5-2\pi \rho\phi^5.
\end{equation}
Under the conformal transformations (\ref{CT1})-(\ref{CT2}) a
symmetric trace-free tensor satisfies
 \begin{equation}
    \bar{\nabla}_{i}\bar{A}^{ij}=\phi^{-10}{\nabla}_i{A}^{ij},
\end{equation}
which simplifies the momentum constraint (\ref{mom}),
\begin{equation}\label{conf mom}
{\nabla}_i {A}^{ij}-\frac{2}{3}\phi^6{\nabla}^jK=8\pi \phi^{10}J^j.
\end{equation} We see immediately that the vacuum (i.e. $\rho=J^j=0$) constraint equations decouple if we
have constant mean curvature, K=const.

The original conformal transverse traceless (CTT) formulation by
York and his later conformal thin sandwich (CTS) formulation have
identical existence and uniqueness properties (see \cite{YCTT}).
They differ in their construction of the tensor ${A}^{ij}$. The CTT
method relies upon tensor splittings to construct a TT tensor. The
CTS formulation is much simpler and bypasses these complications.
There ${A}^{ij}$ takes the form
\begin{equation}
{A}^{ij}=\frac{{\mathbb{L}}\beta^{ij}-{U^{ij}}}{2{N}}
\end{equation} where $U^{ij}$ is the trace-free part of
 the time derivative of the conformal metric, ${N}$ is the conformal lapse related to the
physical lapse $\bar{N}$ by $\bar{N}={N}\phi^6$, $\beta^i$ has a
natural interpretation as the shift vector for the spatial slice,
and $\mathbb{L}$ is the conformal killing operator defined as
${\mathbb{L}}X^{ij}={\nabla}^iX^j+
{\nabla}^jX^i-\frac{2}{3}g^{ij}{\nabla}_kX^k$. The initial data for
the four equation system (\ref{LY}), (\ref{conf mom}) is
$(g_{ij},\;{U}_{ij}, K, {N})$.

The extended conformal thin sandwich (XCTS) system \cite{PY1},
\cite{PY2} extends the CTS system by allowing the slicing
condition to be propagated ($\dot{K}$ is now initial data so that
the evolution equation for $K$ becomes a constraint equation).
This is highly desirable since it is natural to choose $\dot{K}=0$
at quasi-equilibrium (i.e. when ${U}^{ij}=0$), for example, while
it is unclear what choice for the conformal lapse in the standard
CTS formulation should be chosen for this situation. However, this
condition couples the lapse fixing equation $\dot{K}$ to the four
constraint equations and yields a far more complex system with no
constant mean curvature decoupling. The initial data is now in the
Lagrangian form $(g_{ij},\;{U}_{ij};\; K, \dot{K})$. The five
vacuum XCTS equations are
\begin{equation}\label{1}
{\nabla}^2\phi-\frac{{R}(g)\phi}{8}=-\frac{a(\beta)\phi^7}{32\chi^2}+\linebreak
\frac{K^2}{12}\phi^5
\end{equation}
\begin{equation}\label{2}
{\nabla}^2\chi-\frac{{R}(g)\chi}{8}=\frac{7a(\beta)\phi^6}{32\chi}+\frac{5K^2}{12}\chi\phi^4
     -\phi^5({\dot{K}-\beta^i\nabla_{i}K})
\end{equation}
\begin{equation}\label{3}
{\nabla}_i(\frac{\mathbb{L}\beta^{ij}-{U}^{ij}}{2N})-\frac{2\phi^6}{3}\nabla^jK=0
\end{equation}
where $a(\beta)=(\mathbb{L}\beta_{ij}
-U_{ij})(\mathbb{L}\beta^{ij} -{U}^{ij})$ and $\chi=N\phi^7$. This
extended system shares the conformal covariance properties of the
original system by construction since the new equation it couples
to is the physical lapse fixing equation. In the analysis that
follows we restrict our attention to asymptotically flat initial
data.

The non-uniqueness results in \cite{PY1} should not be confused
with the non-uniqueness that results from choosing trivial initial
data in the standard CTT/CTS formulations. In this case we need
only solve (\ref{LY}) which is simply $\nabla^2\phi=0$. We get a
unique regular solution (flat space $\phi=1$) or any number of
singular moment of time symmetry Schwarzschild solutions by
linearity. However, Pfeiffer and York found two regular solutions
for each non-zero wave amplitude considered and their results
suggest that the limiting case of trivial initial data is
identical to the CTT/CTS scenario (as it should be given that the
two systems are essentially equivalent for trivial initial data).
We only consider regular solutions in this work.

In the next section we review the basic uniqueness properties of
the maximal (K=0) CTT/CTS formulation of the constraints. In
section three we focus on the LY equation with unscaled sources.
We assume that a (critical) solution exists whose linearisation
has a kernel and show using Lyapunov-Schmidt theory that a
parabolic branching of solutions. In section four we consider the
non-uniqueness results for the XCTS system. We show that with the
assumption of a 1D kernel, the full XCTS system exhibits a
parabolic branching in all five variables, exactly as was found
numerically in \cite{PY1}. We conclude with a discussion of the
possible implications of these results for evolutions.

\section{The Linear System and Bifurcations}

It is important to understand why the branching features described
in \cite{PY1} and outlined below do not occur in the maximal
CTT/CTS formulation. It is known that the LY equation admits
unique solutions \cite{Cantor} away from trivial initial data. For
the maximal (i.e. decoupled) CTT/CTS systems the existence and
uniqueness of solutions of the LY equation is determined by a
variational inequality for the initial data. A theorem by Cantor
and Brill in \cite{CB}, later corrected by Maxwell in \cite{M},
states that on an asymptotically Euclidean manifold there exists a
positive solution of the LY equation if and only if
$\forall\;f\;\in\;C^{\infty}_0$ the following inequality holds

\begin{equation}\label{yamabe}
inf_{f\;\epsilon\;C^{\infty}_0\;\neq0}\frac{\int\left(|\nabla
f|^2+Rf^2\right)dv}{\|f\|^2_6}>0
\end{equation} where R is the scalar curvature of the conformal
metric and the volume element and inner product are with respect to
this metric. No such theorem is known for the XCTS system. When
studying the $\rho\phi^5$ equation and later the XCTS system we will
emphasise that both systems arise from variation of a non-convex
action so that solutions would correspond to saddle points and are
not expected to be unique.

In non-linear systems the existence of a bifurcating branch of
solutions is indicated by the non-invertibility of the linearized
system. Thus in order to explain the non-uniqueness of solutions
in \cite{PY1} we need to find a background solution whose
linearisation is not invertible.

Given a nonlinear scalar equation $\triangle u=f(x,u)$, such as
(\ref{LY}) with K=0, our first task is to check if it is
linearisation stable, see \cite{OMY}. If $f_{,u}\leq 0$ then the
linearisation has the `¬wrong sign' for use of the maximum
principle to show local uniqueness of solutions. To illustrate
this we recall the following result proven in \cite{CBC}:

On a suitably smooth asymptotically flat background the operator
\begin{equation}\label{iso}
(\nabla^2-f):\;\;H_{k,\delta}\rightarrow H_{k-2,\delta -2}
\end{equation} where $f\geq
0$ and f $\in H_{2,\delta_0}$ with $\delta_o<-2$ (so that f is
continuous and falls off faster than $r^{-2}$, these function
spaces are defined in the appendix) is an isomorphism if
$-1<\delta<0$ and $2\leq k\leq 4$. If we linearise the maximal
CTT/CTS vacuum LY equation about a solution it reduces to this
isomorphic form, so that the implicit function theorem gives us
complete neighbourhoods of nearby solutions (`¬linearisation
stability'). This provides a qualitative explanation for the
uniqueness results found by Pfeiffer and York for the maximal
4-equation CTS system (see Fig. 1 in \cite{PY1}).

In the next section we consider the unscaled source model studied
by York in \cite{Y79} which is not linearisation stable, it's
linearisation has the `¬wrong sign' for application of the maximum
principle so that a kernel solution may exist for sufficiently
large initial data (the function f in (\ref{iso}) is negative). We
demonstrate the existence of a parabolic solution curve similar to
the ones found in \cite{PY1} in a small neighbourhood of such a
critical solution.

\section{Non-Uniqueness for Unscaled Sources}

In \cite{Y79} York considered the case of moment of time symmetry
conformally flat initial data with positive energy density $\rho$ (a
more general analysis of unscaled sources including existence proofs
may be found in \cite{CBIY}). Then the LY equation (\ref{LY}) reads
\begin{equation}\label{rhophi5}
\nabla^2\phi+2\pi\rho\phi^5=0
\end{equation}where $\phi\rightarrow 1$ at spatial infinity. He noted that this
equation is not linearisation stable because the linearisation has
the `¬wrong sign' for use of the maximum principle to prove local
uniqueness.

We now study the form of this non-uniqueness using
Lyapunov-Schmidt (LS) theory. When a linear operator B has a
kernel we are no longer able to invert the equation BV=h(x), with
h(x) given, unless $\int V^*h=0$ where $V^*$ is the kernel of the
adjoint problem i.e. $V^*\in\; Ker B^*=\;coker\;B$. The LS theory
may be regarded as a local extension of these ideas to non-linear
problems. In this method we remove the kernel of the linearisation
from the domain and project the source terms, now non-linear
combinations of the unknown V, onto the image of the
linearisation. In this way a bijective operator $\hat{B}$ is
defined in (\ref{bhat}). The implicit function theorem gives a
unique solution to the mapping between the modified spaces from
which we reconstruct a solution to the original problem from a
Taylor expansion of a real valued function (the LS equation).

We briefly review the LS theory here following closely \cite{VT}
(see also \cite{CH}) whilst keeping in mind the immediate
application to eqn (\ref{rhophi5}). We work in weighted Sobolev
spaces (defined in the appendix) to ensure that the integrals
below are finite, because the Fredholm properties of mappings
between these spaces are well known (see \cite{CBC}, \cite{B}) and
because the implicit function theorem is easily defined for such
spaces. We write the nonlinear equation in functional form
$F(X,\lambda)=0$ where F is a smooth function of the unknown X and
a parameter $\lambda$. F defines a map between Hilbert spaces
(defined in the appendix) $F:X\times\mathbb{R}\rightarrow Y$. To
apply the implicit function theorem F must be at least $C^1$ and
this follows from the multiplication properties of functions in
these spaces. For an example of the implicit function theorem
applied to the CTS formalism with K$\neq0$ between these function
spaces see \cite{CBIY}. Once a solution to the restricted problem
is obtained via the implicit function theorem the remainder of our
analysis is formal.

Our analysis is perturbative and we will \emph{assume} in this
section and the next that there exists a critical solution $X_c$
corresponding to parameter value $\lambda=\lambda_c$. By this we
mean that when we linearise F about $(X_c, \lambda_c)$ we find
that the linear operator $B:=D_XF(X_c,\lambda_c)$ has a kernel
i.e. there is a non-zero function $V_0$ such that $BV_0=0$. We
require that B be fredholm of index zero (i.e. Dim Ker B=Dim coker
B, see appendix) and focus our attention on the case when the
kernel of B is 1D. We perturb $\lambda_c$ by
$\lambda=\lambda_c-\epsilon$ and look for solutions to $F(X,
\lambda)$=0 of the form $X=X_c+V$ where V is a small perturbation.

Clearly $F\;=\;0$ is equivalent to
\begin{equation}\label{bx=r}
BV=R(V,\epsilon)
\end{equation}where $R(V,\epsilon)=BV-F(X,\lambda)$. For equation (\ref{rhophi5}) B is self-adjoint and
(for functions with falloff satisfying $\delta\in(-1,0)$, see
appendix) we have $V_0=V^*_0$ where $V_0^*$ spans the kernel of
$B^*$, the formal adjoint of B. Given $V_0\in H_{k,\delta}$ we
define $z\in C_0^{\infty}$, for simplicity, so that $\int V_0
z=1$. We then split the domain of B as
\begin{equation}\label{dom}
X\subset H_{k,\delta}\;\ni V=\xi V_0+u.
\end{equation}If we define the parameter $\xi=\int
Vzdv$ then $\int uzdv=0$. Thus X is split parallel and
perpendicular to z by this choice of parameter. We take $k\geq 4$
so that if $V\in H_{k,\delta}$ then we also have $V\in C^2$ (see
the appendix).

Similarly we split the range of B as
\begin{equation}\label{ran}
Y\subset H_{k-2,\delta-2}\;\ni R=d z+w
\end{equation}and taking $d=\int RV_0dv$ means $\int
 wV_0dv=0$ and so the Range is split parallel and perpendicular to
 $V_0$.

This splitting allows us to define the following projection
operators
\begin{eqnarray}
  P: &=& X\rightarrow X,\; PX=\xi{V_0}  \\
  Q: &=& Y\rightarrow Y,\; QY=d z.
\end{eqnarray}We now apply separately the projections $1-Q$ and $Q$ to (\ref{bx=r})
to obtain
\begin{equation}\label{bhat}
 \hat{B}u = R(\xi V_0 +u,\epsilon) \\
\end{equation}
\begin{equation}\label{bhatQLS}
  0 =QR(\xi V_0 +u,\epsilon),
\end{equation} where the operator $\hat{B}:(1-P)X\rightarrow (1-Q)Y$ is now
bijective. This means that the linearisation of (\ref{bhat}),
$\hat{B}u=0$, only has the zero solution. The implicit function
theorem then implies that there exists a unique small solution
$u=u(\xi,\epsilon)\in C^2$ to (\ref{bhat}) with $u(0,0)=0$ and
$u_\xi(0,0)=0$ since there are no terms in (\ref{bhat}) linear in
$\xi$.

We now substitute this result into (\ref{bhatQLS}) which yields the
LS equation (see \cite{VT})
\begin{equation}
QR(\xi V_0 +u(\xi V_0,\epsilon),\epsilon)=0.
\end{equation}With $u(\xi V_0,\epsilon)$ a known function given by the implicit function theorem,
 this should be viewed as a real valued equation. It gives gives a
relation between the parameters $\xi=\xi(\epsilon)$.

Since $Q$ is a projection operator we may write the LS equation as
\begin{equation}\label{LS}
d(\xi,\epsilon)= \int_{\mathbb{R}^3} R(\xi V_0 +u(\xi
V_0,\epsilon),\epsilon)V_0dv=0.
\end{equation}i.e. $d=0$ in (\ref{ran}).

At this stage, from the splitting (\ref{dom}), we have a solution to
(\ref{bx=r}) of the form
\begin{equation}
V=\xi(\epsilon) V_0+u(\xi,\epsilon)
\end{equation}
with the particular form of $\xi(\epsilon)$ determined by
(\ref{LS}). We note that for each value of the perturbation
$\epsilon$, $\xi(\epsilon)$ traces a curve along which the
splittings above are valid. If $D_\lambda F(\phi_c,
\lambda_c)\epsilon\neq0$ for some $\epsilon$ then we know that the
set of zeros of $d(\epsilon)$, $d^{-1}(0)$, form a smooth
submanifold (see \cite{CH}).

We now proceed with a formal argument to determine $\xi(\epsilon)$.
Expanding the known $C^2$ function u as a Taylor series in the
dependent parameter $\xi$ and the independent parameter $\epsilon$
we have
\begin{equation}
u(\xi,\epsilon)=\epsilon u_{\epsilon}(0,0)+O(2):=\epsilon u^*+O(2)
\end{equation}
since $u_{\xi}(0,0)\neq 0$ is incompatible with (\ref{bhat}) due to
the absence of linear terms in V in R (where subscripts denote
partial derivatives and O(2) denotes second order terms in $\xi,
\epsilon$). Now our solution to (\ref{bx=r}) is
\begin{equation}\label{LS expn}
V=\xi(\epsilon) V_0+\epsilon u^*+O(2).
\end{equation}

The LS equation may now be written in the form
\begin{equation}\label{LS eqn}
L_{20}+\xi\epsilon L_{11}+\epsilon L_{01}+...=0
\end{equation}
where $L_{mn}=\int V_0R_{mn}dv$ and $R_{mn}$ denotes the mth order
term in $\xi$ and the nth order term in $\epsilon$ resulting from
substitution of the solution (\ref{LS expn}) into $R(V, \epsilon)$.
This fixes $\xi=\xi(\epsilon)$.

In \cite{VT} Vainberg and Trenogin prove that the small solutions
(i.e. where $\xi(\epsilon)\rightarrow 0$ as $\epsilon\rightarrow 0$)
of the LS equation are in 1-1 correspondence with the small
solutions of $F(X, \lambda)=0$.

We now show that branching of the type found for the XCTS system
in \cite{PY1} is actually a generic property of solutions of the
standard LY equation with an unscaled source (this was shown
recently in the special case of the constant density star in
\cite{BOMP}). We examine the local behavior of the solutions to
this equation at a critical point of the linearisation as the
unscaled source term is varied. We multiply $\rho$ in
(\ref{rhophi5}) by a positive parameter $\lambda$ and seek a
continuous family of solutions to the LY equation
\begin{equation}\label{unsc}
\nabla^2\phi+2\pi\rho\lambda\phi^5=0, \;\;\phi\rightarrow 1\;\;
as\;\; r\rightarrow\infty
\end{equation}on a fixed flat background. If $\rho$ is a suitably smooth compactly supported function and we
seek solutions $\phi$ such that $\phi-1\in\;H_{k,\delta}$ where
$\delta\in(-1,0)$ then the integral relations above will be finite.

For $\lambda=0$ the only regular solution satisfying the boundary
conditions is $\phi\equiv1$. As we increase $\lambda$ we find that
the linearised homogenous equation
\begin{equation}\label{kernel}
\nabla^2V_0+10\pi\rho\lambda\phi^4V_0=0
\end{equation}has only the trivial solution $V_0\equiv 0$. The
implicit function theorem then tells us that (\ref{unsc}) has a
smooth sequence of solutions, $\phi(\lambda)$, with $\phi(0)=1$
and the maximum principle tells us that $\phi(\lambda)\geq 1$. We
now assume that there exists a critical solution $\phi_c$ at
$\lambda=\lambda_c$, whose linearisation has a kernel $V_0$,
corresponding to the lowest eigenstate of the Schrodinger-type
equation (\ref{kernel}) . We know that this eigenstate is unique
up to scaling and has no nodes, so we may take $V_0>0$.

We now expand (\ref{unsc}) about this critical value where the
linearisation, (\ref{kernel}), has a kernel. Taking

$\phi:=\phi_c+V$ with $\phi_c>0$ and $\lambda:=\lambda_c-\epsilon$
we find

\begin{equation}
\nabla^2(\phi_c+V)+2\pi\rho(\lambda_c-\epsilon)(\phi_c+V)^5=0
\end{equation}which gives

\begin{eqnarray}\label{B}
BV&=&(\nabla^2+10\pi\lambda_c\phi^4_c\rho)V \\
&=&2\pi\epsilon\rho\phi^5_c-20\pi\lambda_c\rho\phi^3_c V^2 +
10\pi\rho\phi^4_c\epsilon V + 7\; other\; terms.
\end{eqnarray}The LS equation (\ref{LS}) for this problem is
\begin{equation}\label{QRrho}
\int_{\mathbb{R}^3}
V_0\left(2\pi\rho\phi^5_c+10\pi\rho\phi^4_c\epsilon
V-20\pi\lambda_c\rho\phi^3_cV^2+...\right)dv=0.
\end{equation}If we substitute the solution V in the form (\ref{LS expn})
into (\ref{QRrho}) we obtain the coefficients $\L_{mn}$:
\begin{eqnarray}\label{lij}
  L_{01} &=&\int_{\mathbb{R}^3} V_0(2\pi\rho\phi^5_c)dv >0\\
  L_{11} &=&\int_{\mathbb{R}^3}
  V_0(10\pi\rho\phi^4_cV_{0}-40\pi\lambda_c\rho\phi^3_cV_{0}u^*)dv\\
  L_{20} &=&\int_{\mathbb{R}^3} V_0(-20\pi\lambda_c\rho\phi^3_cV_{0}^2)dv<0,
\end{eqnarray}and so on.

By choosing $\xi$ and $\epsilon$ small enough we may truncate the LS
branching equation (\ref{LS eqn}) at any order. In particular we can
write

\begin{equation}
\xi^2L_{20}+\xi\epsilon|L_{11}| + \epsilon L_{01}\approx0.
\end{equation}

Solving the quadratic equation for $\xi$ we find,
\begin{equation}\label{bifurc}
\xi=\pm\left(\frac{L_{01}}{|L_{20}|}\epsilon\right)^{\frac{1}{2}}+o(\sqrt{\epsilon}).
\end{equation} where
$\frac{o(\sqrt{\epsilon})}{\sqrt{\epsilon}}\rightarrow 0$ as
$\epsilon\rightarrow 0$. This tells us that to lowest order and
provided $L_{01}\neq 0$ that we may ignore the contribution of
$L_{11}$.

Therefore, in a small neighbourhood of the critical solution
$\phi_c$ we find that as the parameter $\epsilon$ is varied that the
conformal factor traces a parabola in the solution space
\begin{equation}\label{phi para}
\phi=\phi_c
\pm\left(\frac{L_{01}}{|L_{20}|}\epsilon\right)^{\frac{1}{2}}V_0+O(\sqrt{\epsilon}).
\end{equation}

If $L_{01}= 0$ then $2\pi\rho\phi^5_c\;\in$ Image(B) and a
qualitatively different situation arises. The LS equation then
tells us that there is more than one branch of solutions passing
through $\lambda=\lambda_c$ provided not all $L_{ij}\neq 0$. In
this case the zeros of the bifurcation equation (\ref{QRrho}) do
not form a smooth submanifold. Note that this is not the behaviour
observed in \cite{PY1} where only a single smooth (parabolic)
curve of solutions is found. This phenomenon was observed recently
in a numerical study of non-unique solutions to (\ref{rhophi5})
corresponding to a constant density star, \cite{BOMPpc}. Further
details can be found in \cite{CH}.

As noted above, specialising (\ref{rhophi5}) to the constant
density star, upper and lower branches of solutions were found in
\cite{BOMP} and an explicit form for the critical and kernel
solutions were given. Generalising from $\rho=const$ and compactly
supported to just $\rho>0$ and compactly supported, only the lower
branch of solutions was given and a parabolic approach to the
critical solution was found in \cite{BOMP}. The existence of an
upper branch was conjectured and is easily determined using the LS
methods outlined above which give the required
$\pm\sqrt{\epsilon}$ behavior in a small neighbourhood of the
critical point.

In \cite{Y79} York notes that by specifying a conformal energy
density $\hat{\rho}$ related to the physical energy density by
$\rho=\hat{\rho}\phi^{-s}$ where $s>5$, we transform the
linearisation of (\ref{unsc}) into an isomorphic form yielding
unique solutions. Therefore non-uniqueness results from a poor
choice of conformal scaling in this case. We argue in the next
section that the coupling of the $\dot{K}$ evolution equation to the
four constraint equations in the XCTS system similarly introduces an
undesirable scaling of the variables leading to non-uniqueness as
above.

\section{The XCTS System}

As in the case of unscaled sources above, to apply the LS method
we must first assume the existence of a critical solution about
which the linearisation of this system has a (vector) kernel.

We note here that the XCTS system arises from the variation of an
action given by
\begin{equation}\label{action}
\mathcal{I}(u)=\int_M\left(\nabla\chi.\space\nabla\phi+\frac{R\phi\chi}{8}+
\frac{a(\beta)\phi^7}{32\chi}+\frac{K^2}{12}\chi\phi^5-
\frac{\phi^6}{6}({\dot{K}-\beta^i\nabla_{i}K})\right)dv.
\end{equation}

We may also include scaled versions of the energy density and the
current density. This allows us to examine the existence and
uniqueness properties of the XCTS system from a variational
perspective.

First we investigate the necessary conditions for the functional
(\ref{action}) to have a minimum. We must have that the functional
is bounded below on an appropriately defined domain of admissible
functions $\Pi$ which will be some Sobolev space $H_{k,\delta}$
subject to the constraints $N>0,\;\phi>0$.

Provided such a domain has been defined, we proceed to the
calculation of the first and second variations of $\mathcal{I}$. If
we define $J(s)=\mathcal{I}(u+sv)$, where $v\in \Pi$, a stationary
value of $\mathcal{I}$ is given by $J'(0)=0$ where $'$ denotes
differentiation with respect to s. A simple calculation reveals that
the XCTS system corresponds to a stationary point of this
functional.

It is straightforward to show that the second variation of
(\ref{action}) is not of definite sign. The following inequality is
used in \cite{G} as the basic assumption on the integrand F to apply
direct methods in the calculus of variations;
\begin{equation}
F_{P^i_{\alpha}P^j_{\beta}}(x,u,\nabla
u)\xi^i_{\alpha}\xi^j_{\beta}>0\;\;\;
\end{equation}for all rank one matrices $\xi^i_\alpha$ where i=1-5,
$\alpha$=1-3 and $P^i_{\alpha}$ denote the usual derivatives in the
Euler-Lagrange equations. This condition, called `¬strong
ellipticity' in \cite{G}, is the requirement that $F(x, u, p)$ be
convex with respect to p ($p=\nabla u\;\in\;
\mathbb{R}^{5\times3}$). The XCTS Lagrangian (\ref{action}) fails
this criterion; it is convex in $\mathbb{L\beta}$ but, due to the
mixed term $\nabla\chi\nabla\phi$, F is not convex in $\nabla\phi$
or in $\nabla\chi$.

This lack of convexity means we cannot expect stationary points, if
they exist, to be unique. An analogous situation occurs frequently
in non-linear elasticity where non-convex functionals are necessary
in order to model buckling equilibrium configurations of materials
which are known physically to be non-unique, i.e. an input stress
can lead to numerous buckled states. This argument and the analogy
to the unscaled source equation (\ref{rhophi5}) lends support to our
assumption below that the linearised XCTS system develops a kernel
for sufficiently large initial data.

\subsection{Non-uniqueness in the XCTS system}

In this section we assume that the XCTS system (\ref{1})-(\ref{3})
has a critical solution $\overrightarrow{X}_c$ and that the
linearisation about this solution has a one dimensional kernel
$\overrightarrow{V}_0$. To apply the LS method we must also check
that the formally adjoint system has a kernel of equal dimension.
We know from \cite{CBC} that for sufficiently smooth initial data
with very general falloff conditions at spatial infinity that the
kernel of this linearised system is finite dimensional and that it
has a closed range. We show in the appendix that the linear system
is actually Fredholm with an index of zero (so that dim Ker=Dim
coKer) between suitably defined Sobolev spaces. Assuming now that
the kernel is one dimensional, we have satisfied the requirements
to implement the LS theory outlined above.

We introduce a parameter $\lambda$ to allow us to continuously
vary the conformal initial data $g_{ij}$ and $U^{ij}$ in the
conformal background(in \cite{PY1} the parameter `¬A' was used
which corresponded to the amplitude of a Teukolsky wave in the
conformal background). The 1-parameter family of initial data
considered in \cite{PY1} is
\begin{eqnarray}
g_{ij}&=&\delta_{ij}+\lambda h_{ij}\\
U_{ij}&=&\lambda\dot{\hat{h}}_{ij}.
\end{eqnarray} where $\hat{h}_{ij}$ is the tracefree part of the metric
perturbation (these tensors are given explicitly in \cite{PY1}).
This corresponds to a gravitational wave perturbation of flat
space with a gaussian wave profile. The perturbation $h_{ij}$
decays exponentially with distance so the metric is asymptotically
flat. (Our construction is valid for systems with much weaker
power-law falloff). Furthermore, due to the fast falloff, we know
that this conformal metric will not contribute to the ADM energy
of the physical solution.

In (\ref{1})-(\ref{3}) we denote the dependence of the initial data
on $\lambda$ by
\begin{eqnarray*}
  \nabla\hookrightarrow\nabla_\lambda, & U^{ij}\hookrightarrow\lambda U^{ij}, &R\hookrightarrow R(\lambda). \\
\end{eqnarray*}
When $\lambda=0$ the XCTS equations (\ref{1})-(\ref{3}) decouple due
to the non-existence of conformal killing vectors that vanish at
infinity (where we have the constraint on the conformal lapse that
$N>0$), and we obtain flat space as the unique regular solution.
Applying the implicit function theorem then gives a local curve of
solutions parameterised by $\lambda$. We assume that there exists a
critical solution $\overrightarrow{X}_c=(\phi_c,
 \chi_c,\;\beta^i_c)$ occurring at $\lambda=\lambda_c$. At
this point we apply the LS method to continue the curve through
$\lambda_c$.

 Below we will, as in the previous example,
 perturb the system about this critical solution. It is convenient
 to absorb the critical shift vector into
 $U^{ij}$ by a gauge transformation.
We perturb the XCTS system (\ref{1})-(\ref{3}) at the critical
solution $\overrightarrow{X}_c=(\phi_c,\chi_c,0,0,0)$ according to
\begin{eqnarray*}
\lambda&=&\lambda_c-\epsilon\\
\phi&=&\phi_c+\phi_1 \\
\chi&=&\chi_c+\chi_1 \\
\beta^i&=&0+\beta^i_1.
\end{eqnarray*}

We expand the background scalar curvature in a Taylor series about
$\lambda_c$ so that
 \begin{equation}\label{r expn}
 R(\lambda)=R(\lambda_c)-\epsilon R'(\lambda_c)+...
 \end{equation}

The linear terms in the expansion give the following inhomogenous
system
\begin{eqnarray*}
  \Delta \phi_1
  -\frac{1}{8}R(\lambda_c)\phi_1+\frac{\phi^7_c}{32\chi^2_c}
  \left(-2U.\mathbb{L}\beta_1-(2\frac{\chi_1}{\chi_c}-7\frac{\phi_1}{\phi_c})U.U\right)&=&
   \epsilon \frac{R'(\lambda_c)}{8}\phi_c +\epsilon\Gamma\\
\Delta \chi_1
-\frac{1}{8}R(\lambda_c)\chi_1+7\frac{\phi^6_c}{32\chi_c}
\left(2U.\mathbb{L}\beta_1-(6\frac{\phi_1}{\phi_c}-\frac{\chi_1}{\chi_c})U.U\right)&=&
 \epsilon \frac{R'(\lambda_c)}{8}\phi_c  + \epsilon\Gamma \\
\nabla_i\left(\frac{\phi_c^7\mathbb{L}\overline{\beta}^{ij}_1}{\chi_c}-\frac{\phi^7_c}{\chi_c}U^{ij}
(7\frac{\phi_1}{\phi_c}-\frac{\chi_1}{\chi_c})\right)&=& \epsilon
\Gamma^j
\end{eqnarray*}where terms arising from variation of the connection
are given the generic symbol $\Gamma$. Due to the form of the
conformal initial data we are considering we know that these
connection terms decay exponentially with distance.

The XCTS system (\ref{1})-(\ref{3}) will be referred to here as
$F(\overrightarrow{X},\lambda)=0$. As before, F describes a
mapping between Hilbert spaces $F:X\times\mathbb{R}\rightarrow Y$.
We assume the existence of a 1D kernel $\overrightarrow{V}_0$ and
define the dual object $\overrightarrow{Z}\in C_0^{\infty}$ so
that
$\int_{\mathbb{R}^3}\overrightarrow{V}_0.\overrightarrow{Z}dv=1$
where the inner product denotes multiplication of a five component
row vector with a 5 component column vector and the volume element
is taken with respect to the conformal metric $g_{ij}$. Then we
may split the domain X as five copies of the one defined in the
scalar case (\ref{dom}) and similarly the range corresponds to
five copies of (\ref{ran}). Likewise our relation for $\xi$ in the
scalar case, namely $\xi=\int{Vz}dv$, becomes
\begin{equation}
\xi=\int_{\mathbb{R}^3} \overrightarrow{V}.\overrightarrow{Z}dv.
\end{equation} The critical solution satisfies
$F(\overrightarrow{X}_c,\lambda_c)=0$ and $\overrightarrow{V}_0$
satisfies
$$B\overrightarrow{V}_0=D_XF(\overrightarrow{X}_c,\lambda_c)\overrightarrow{V}_0=0.$$
The LS machinery developed for the scalar equation (\ref{rhophi5})
naturally generalises to systems of elliptic equations. As before we
have that $F=0$ is equivalent to
\begin{equation}
B\overrightarrow{V}_0:=D_X
F(\overrightarrow{X}_c,\lambda_c)\overrightarrow{V}_0=\overrightarrow{R}(V,\epsilon).
\end{equation}

Following the same procedure as (\ref{bhat})-(\ref{bhatQLS}),
restricting the domain and range of B by applying the projections
1-Q and Q to $\overrightarrow{R}$ gives us
\begin{equation}\label{bhatsys}
\hat{B}\overrightarrow{u}=\overrightarrow{R}(\xi
\overrightarrow{V}_0+\overrightarrow{u}, \epsilon)
\end{equation}
\begin{equation}\label{proj sys}
Q\overrightarrow{R}(\xi \overrightarrow{V}_0+\overrightarrow{u},
\epsilon)=0
\end{equation}
where Q denotes the projection onto $\overrightarrow{Z}$ as before.
The implicit function theorem is also valid for nonlinear systems
(see for example \cite{CBIY}). The linearisation of (\ref{bhatsys})
now reads $\hat{B}\overrightarrow{u}=0$. Since $\hat{B}$ is now an
isomorphism, we obtain a unique small solution
$\overrightarrow{u}(\xi,\epsilon)$ to (\ref{bhatsys}) with
$\overrightarrow{u}(0,0)=\overrightarrow{u}_{\xi}(0,0)=0$. We
substitute this into (\ref{proj sys}) to obtain the orthogonality
relation that defines the curve $\xi=\xi(\epsilon)$. It is worth
pointing out that, since the linear system is fredholm with index
zero and the kernel is assumed to be one dimensional, we still have
just one equation, (\ref{proj sys}), for one unknown $\xi$ (though
the numerical value of $\xi$ now depends on all five unknowns).

At this point we have a (vector) solution of the form
\begin{eqnarray}\label{sys expn}
\overrightarrow{V}&=&\xi(\epsilon)\overrightarrow{V}_0+\overrightarrow{u}(\xi,
\epsilon)\\&=&\xi(\epsilon)\overrightarrow{V}_0+\epsilon\overrightarrow{u}^*
+O(2),
\end{eqnarray}
just as before (see (\ref{LS expn})). Equation (\ref{proj sys})
reads
\begin{equation}
0=\int_{\mathbb{R}^3}\overrightarrow{V^*_0}.(\overrightarrow{R})dv=\int(V_0^{*\phi},
V_0^{*\chi},V_0^{*\beta1} , V_0^{*\beta 2}, V_0^{*\beta
3}).\left(%
\begin{array}{c}
   R^{\phi}\\
   R^{\chi} \\
   R^{\beta_1} \\
   R^{\beta_2} \\
  R^{\beta_3} \\
\end{array}%
\right)dv
\end{equation}where $R^{\phi}$ corresponds to the first order in $\epsilon$
background term, $D_{\lambda}F(X_c, \lambda_c)\epsilon$, and
nonlinear terms arising on the RHS of the Hamiltonian constraint.
Note that in the appendix we have shown that $$(V_0^{*\phi},
V_0^{*\chi},\frac{1}{8}V_0^{*\beta1} , \frac{1}{8}V_0^{*\beta 2},
\frac{1}{8}V_0^{*\beta 3})=(V_0^{\chi}, V_0^{\phi},V_0^{\beta1} ,
V_0^{\beta 2}, V_0^{\beta 3}).$$ Likewise the LS coefficients are
given by
\begin{equation}
L_{ij}=\int_{\mathbb{R}^3}
\overrightarrow{V_0}^*.\overrightarrow{R_{ij}}dv.
\end{equation}where $\overrightarrow{R}_{ij}$ denotes the ith order term in $\xi$
and the jth order term in $\epsilon$ resulting from substitution of
the solution (\ref{sys expn}) into $\overrightarrow{R}$.

The source terms on the RHS of (\ref{bhatsys}) are of the form
\begin{equation}\label{sys-nonlin}
(\overrightarrow{R})=\left(\begin{array}{c}
   R^{\phi} \\
   R^{\chi} \\
   R^{\beta_1} \\
   R^{\beta_2} \\
   R^{\beta_3} \\
\end{array}%
\right)=\left(%
\begin{array}{c}
  -\frac{1}{8}\epsilon R'(\lambda_c)+\epsilon\Gamma +T_1(\phi,\chi, L\beta)+... \\
  -\frac{1}{8}\epsilon R'f
  (\lambda_c)+\epsilon \Gamma+T_2(\phi,\chi, L\beta)+...\\
  \epsilon\Gamma+T_3(\phi,\chi, L\beta)+... \\
  \epsilon\Gamma +T_4(\phi,\chi, L\beta)+...\\
  \epsilon\Gamma +T_5(\phi,\chi, L\beta)+...\\
\end{array}%
\right)
\end{equation}

where $T_i$ represents the quadratic combinations of all variables
arising from the perturbation. From the form of the solution
(\ref{sys expn}) we see that the $T_i$ terms yield terms quadratic
in $\xi$ and $\epsilon$ and mixed terms proportional to
$\xi\;\epsilon$.

Clearly
\begin{equation}\label{l01}
L_{01}=\int_{\mathbb{R}^3} \left((V_0^\phi+V_0^\chi)\frac{1}{8}
R'(\lambda_c)-\overrightarrow{V_0}.\Gamma\right) dv.
\end{equation}

In the scalar model (\ref{rhophi5}) we worked on a fixed flat
background so that the connection wasn't varied. We also knew that
the first eigenfunction of the linearised equation had no nodes.
When dealing with a coupled system of equations we lose this
property. To obtain the results of \cite{PY1} we must assume that
$L_{01}\neq0$, which means that only a single curve of solutions
passes through the critical point ($d^{-1}(0)$ is a smooth
submanifold).

The next possibly non-zero terms are $L_{11}$ and $L_{20}$. If
$L_{01}\neq 0$ and $L_{20}\neq0$ then we may truncate our series
at quadratic order and ignore the contribution of $L_{11}$. The
complexity of the combinations of the quadratic terms in
(\ref{sys-nonlin}) do not yield easily to analysis. We prefer to
emphasise that it is extremely unlikely that our choice of initial
data could lead to the cancellation of all terms at this order.
Furthermore, since the kernel solution is removed by the LS method
we know that the operator $\hat{B}$ is an isomorphism which
implies that all the $L_{ij}$'s are finite for small $\xi$ and
$\epsilon$. We conclude that

\begin{equation}
\xi^2L_{20}+\epsilon L_{01}\approx 0.
\end{equation}

In principle the numbers $L_{01}$, $L_{20}$ can be determined for
the choice of initial data that yields the kernel solution
$\overrightarrow{V}_0$ (c.f. (\ref{lij})). However, by choosing the
sign of $\epsilon$ so that $\frac{L_{01}\epsilon}{L_{20}}<0$ we know
that there is a parabolic branching of all five variables
\begin{equation}\label{sys parab}
\overrightarrow{X}\approx\overrightarrow{X}_c\pm\left(\frac{|L_{01}\epsilon|}{|L_{20}|}\right)^{\frac{1}{2}}\overrightarrow{V}_0.
\end{equation}(c.f. equation (7) in \cite{PY1}). The parabola in \cite{PY1} corresponds to the case where
$\frac{L_{01}}{L_{20}}<0$ so that $\epsilon$ must be positive (i.e.
there are no solutions with $\lambda>\lambda_c$).

To recap, we list the assumptions that were made to derive
(\ref{sys parab}). The first assumption was that for sufficiently
large $\lambda$ the curve of exact solutions (found numerically in
\cite{PY1}) reaches a critical solution (an exact solution whose
linearisation has a 1-D kernel $\overrightarrow{V}_0$) and that
none of the components of $\overrightarrow{V}_0$ vanishes
identically. If a component of $\overrightarrow{V}_0$ was
identically zero then the lowest order term in the expansion
(\ref{sys expn}) of this component would not vary like
$\pm\sqrt{\epsilon}$ in the vicinity of $\lambda_c$, contrary to
the results in \cite{PY1}.

Secondly, we assumed that $L_{01}\neq 0$ and $L_{20}\neq 0$. The
very general form of the integral (\ref{l01}) and the
corresponding integral for $L_{20}$ suggests that this is by far
the likeliest outcome. This guarantees a parabolic curve such as
(\ref{sys parab}). The solution curves found numerically by
Pfeiffer and York correspond to $\frac{L_{01}}{L_{20}}<0$.

\section{Implications for evolutions}
We now consider the likely implications of these results for
evolutions of the constraints. Recall that the Bianchi identities
tell us that if the constraints are satisfied on the initial slice
then they are satisfied on all slices of the foliation. However in
numerical evolution schemes constraint violation off the initial
data hypersurface remains a serious problem. To control constraint
errors one may choose to solve the constraints on each slice of
the foliation (constrained evolution).

The Bianchi identities also make some of the Einstein equations
redundant. Given a solution of the constraints $(\bar{g}=\phi^4 g,
K)$, we may choose to evolve the conformal factor according to the
evolution equation for the metric. For example, evolving initial
data from the XCTS formulation naturally gives
\begin{equation}\label{phidot}
(\partial_t-\beta^i\partial_i){\log\phi}=\frac{1}{6}(-N\phi^6K+\nabla_i\beta^i)
\end{equation}
where N is the conformal lapse and derivatives are with respect to
the conformal metric. Whereas (\ref{phidot}) selects just one
solution of the constraints to evolve, the constrained evolution
scheme could possibly lead to a solution jumping between branches
during an evolution.

From equation (\ref{phidot}) we see that if we choose maximal
slicing with zero shift then $\dot{\phi}=0$ and so a constrained
evolution of this data should yield the initial conformal factor
throughout the evolution. This provides a possible test of whether
a constrained evolution has caused $\phi$ to stray from the
correct branch.

The example (\ref{rhophi5}) serves to illustrate the need to
choose an appropriate scaling of the extrinsic curvature. The
standard scaling of the LY formulation in (\ref{CT2}) not only
simplifies the momentum constraint but also removes the
linearisation instability. Many axisymmetric evolution schemes
(see \cite{GD}, \cite{Choptuik}) do not scale the momentum. This
need not cause problems on the initial slice if moment of time
symmetry data is chosen. However, constrained evolution of this
initial data could prove problematic as later time slices are
susceptible to linearisation instability and non-uniqueness as
noted in \cite{Rinne}. Also, the conformal scaling in the
Hamiltonian constraint in the BSSN formulation is analogous to
(\ref{rhophi5}). This is not necessarily a problem if an initial
data set is transformed into this formalism and then evolved using
a free evolution such as (\ref{phidot}).

\section{Conclusion}

In \cite{PY1} the solution curves for initial data in CTS and XCTS
were compared. As expected the CTS system had unique solutions for
$\lambda<\lambda_{crit}$ and no solutions for
$\lambda>\lambda_{crit}$. With identical initial data $(g_{ij},
U_{ij})$, but with $\dot{K}=0$ replacing N=1 the solution curve
for the XCTS system is parabolic and turns back upon itself at the
(much smaller) critical wave amplitude. We have shown that this is
consistent with the system developing a kernel solution of its
linearisation and being dominated by a quadratic non-linearity
there.

It is worth emphasising that all the non-uniqueness results in
this work are local in nature, i.e. the multiplicity of solutions
is confined to a small neighbourhood of the critical solution. We
have argued that the parabolic branching behaviour in \cite{PY1}
is the simplest example of branching phenomena that can arise in
non-linear elliptic PDEs. However, the global nature of the
parabolic branches in \cite{PY1} is surprising.

The non-uniqueness properties described above are generic in the
sense that they do not depend on the form of the (Fredholm)
linearised operator B nor on the form of the forcing terms. If B
has a 1D kernel and there are quadratic nonlinearities then a
parabolic solution curve is to be expected (provided
$L_{01},\;L_{20}\neq0$).  With higher order non-linearities or
with multiple parameters $\lambda_i$ more complicated solution
curves arise. It is therefore natural to expect non-uniqueness (in
a neighbourhood of the critical solution) in any non-linear
elliptic system with a non-trivial kernel.

Whilst our results using LS theory mirror the numerical results
found in \cite{PY1}, it is important to remember that this
non-uniqueness is removed when the wave amplitude $\lambda$ picks
up a spatial dependence when we map from the conformal to the
physical space. In the original paper it was noted that in the
physical space the wave amplitude was not $\lambda$ but rather
$\phi^4\lambda$ and that using instead $\lambda$max$\phi^4$ as our
wave amplitude parameter leads to a solution curve with no
turn-around point, solutions are seen to be unique for this
"parameter" (see fig.4 in \cite{PY1}). We may draw a parallel to
the scalar equation (\ref{rhophi5}) for which York noted that
choosing an appropriate conformal scaling for $\rho$ leads to
linearisation stability and unique solutions but at the
considerable cost of losing control over a physical quantity, the
energy density $\rho$.

The XCTS system has been very successful in modelling Black holes
using the puncture technique and trivial initial data
corresponding to quasi-equilibrium i.e. $(U_{ij}=K=\dot{K}=0)$ for
which the system considerably simplifies (see, for example,
\cite{Hannam} where the value of the shift at the "puncture" was
specified to prevent it vanishing everywhere).

When we studied the XCTS system in this work we considered only
Dirichlet boundary conditions on an asymptotically flat manifold
containing gravitational waves. When modelling black holes an
alternative to the puncture technique is to excise a region
corresponding to an apparent horizon. Much analytical work has
been done to guarantee unique solutions satisfying apparent
horizon boundary conditions in the standard CTT formalism. A
recent numerical study of the generalisation of these methods to
the XCTS system revealed many difficulties with well-posedness
(see \cite{jar} and references therein) as should be expected from
the results outlined here. These results add weight to our
reservations regarding the possible dangers of constrained
evolution using ill-posed formulations of the constraints.

When modelling data with $U_{ij}\neq 0$, or when using apparent
horizon boundary conditions it is clear that the XCTS system
becomes much more complicated and it may prove more beneficial to
revert to the standard and much simpler four equation CTS
formalism.

\section*{Acknowledgment}\noindent

It is a pleasure to thank my PhD advisor Niall O Murchadha for his
advice and kindness. I also wish to thank Thomas Baumgarte, Sergio
Dain, Harald Pfeiffer and in particular Edward Malec who pointed
out the classic reference \cite{VT}. Sincere thanks also to Oliver
Rinne for discussions on evolving the constraints and to the
referee for helpful suggestions on clarifying this work. This work
was funded by an IRCSET embark scholarship.

\section*{Appendix: Function spaces and the linearised system}

We now define the function spaces used in the text. Our analysis has
been restricted to asymptotically flat manifolds $(\mathbb{R}^3,
\bar{g}_{ij})$ where $\bar{g}-\delta$ satisfies the falloff
conditions below.

We use weighted Sobolev spaces of tensors with norm defined as (see
\cite{B})
\begin{equation}
\|W\|^2_{H_{k,\delta}(\mathbb{R}^3)}=\sum^k_{m=0}\int_{\mathbb{R}^3}|\nabla^mW|^2\sigma^{-2(\delta-m)-3}dv
\end{equation} where $\sigma=(1+r^2)^{\frac{1}{2}}$, $r=|x|$ is the
Euclidean distance function and the volume form and covariant
derivatives are with respect to the Euclidean metric $\delta$.

The conformal metric considered in \cite{PY1} was a perturbation
to flat space consisting of a quadrupole gravitational wave with a
gaussian profile. The metric therefore had an exponential falloff
with distance. When we conformally map to a solution of the
constraints we know that this background metric will make no
contribution to the ADM energy.

Through the conformal method we aim to construct a Riemannian
asymptotically flat 3 metric $\bar{g}$ where
\begin{eqnarray}
  \bar{g}_{ij}-\delta_{ij} &\in& H_{k,\delta} \\
  \bar{U}_{ij} &\in & H_{k-1,\delta-1}
\end{eqnarray} and we seek
solutions $\phi-1,\;1-\chi,\;\beta^i\;\in\;H_{k,\delta}$ where
$k\geq 4$ and $\delta\in (-1,\;0)$. k is the number of times a
tensor is weakly differentiable. Note that with $k\geq 4$ the
metric is $C^2$ (we lose $3/2$ degrees of differentiability in
passing from weak to strong differentiability). If v $\in
H_{k,\delta}$ with $k>3/2$ then $|v(x)|=o(r^\delta)$ as r tends to
infinity, see \cite{B}.

The linearised XCTS system is a mapping
\begin{equation}
B:\;H_{k,\delta}\longrightarrow\;H_{k-2,\delta-2}.
\end{equation}
and was given after equation (\ref{r expn}) above.

The formally adjoint system is defined for all u $\in\;H_{k,\delta}$
according to
\begin{equation}
<Bu,v>=\int vBu=\int uB^*v
\end{equation}where inner products denote a product of five component row and
column vectors, the volume form is with respect to the background
metric and $v\in H_{k-2,-1-\delta}$. Thus $B^*$ is a mapping
\begin{equation}
B^*:\;H_{k-2,-1-\delta}\longrightarrow\;H_{k-4,-3-\delta}.
\end{equation}given by
\begin{eqnarray}
 \Delta \hat{\phi}_1
  -\frac{1}{8}R(\lambda_c)\hat{\phi}_1+7\frac{{\phi}^6_0}{32\chi_0}
  \left(16U.\mathbb{L}\hat{\beta}_1+(-6\frac{\hat{\chi}_1}{\phi_0}+\frac{\hat{\phi}_1}{\chi_0})U.U\right)&=&0\\
\Delta \hat{\chi}_1
  -\frac{1}{8}R(\lambda_c)\hat{\chi}_1+\frac{\phi^7_0}{32\chi^2_0}
  \left(-16U.\mathbb{L}\hat{\beta}_1+(7\frac{\hat{\chi}_1}{\phi_0}-2\frac{\hat{\phi}_1}{\chi_0})U.U\right)&=&0\\
\nabla_i\left(\frac{\mathbb{L}\hat{\beta}^{ij}_1\phi^7_0}{\chi_0}+\frac{\phi^7_0}{8\chi_0}U^{ij}
(\frac{\hat{\phi}_1}{\chi_0}-7\frac{\hat{\chi}_1}{\phi_0})\right)&=&0.
\end{eqnarray}
We now outline the Fredholm properties of $B$. Linear elliptic
systems on asymptotically Euclidean manifolds of the form
$Bu=\sum^m_{k=0}a_kD^ku$ were studied in \cite{CBC}. Clearly the
linearised XCTS system is elliptic. Our initial data satisfies
hypothesis one of that work regarding smoothness and falloff. They
proved that such a system has a finite dimensional kernel and a
closed range. This says that B is semi-Fredholm so that the domain
of B splits as $H_{k,\delta}=kerB + W$ where W is closed and B is
injective on W. However, this is not good enough for the application
we have in mind. We need B to be Fredholm with an index of zero (so
that the dimension of the cokernel is equal to that of the kernel of
B).

The domain of B is contained in $ H_{k,\delta}$ such that $\delta
\in(-1,\;0)$. This implies that $-1-\delta\in(-1,\;0)$ so that the
domain of $B^*$ is the same as that of B. From a theorem in
\cite{CBC} we know that each system has a finite dimensional kernel
and a closed range. At first sight it is not clear that the kernels
of B and $B^*$ are related. However, inspection of the systems
reveals that if $\overrightarrow{V}_0=(\phi_1, \chi_1,
\beta^i_1)\in\;ker\;B$ then $\overrightarrow{V}^*_0:=(\chi_1,
\phi_1, \frac{1}{8}\beta^i_1)\in\;ker\;B^*$ and so the systems are
identical under this relabelling of variables. Therefore the systems
have kernels of equal finite dimension so that B is Fredholm with an
index of zero. This is the basic requirement for the use of the LS
methods outlined above. If the linearised system is Fredholm of
index zero then we know that the LS relation QR=0 gives n equations
in n unknowns $\xi_i$ where n is the dimension of the kernel. If Dim
Ker is 1 then we have a single equation for $\xi$ in terms of
$\epsilon$.


\begin{thebibliography}{99}
\bibitem{RW}These ideas will be expanded in future work with
Oliver Rinne.
\end{thebibliography}
\end{document}